\documentclass[journal = jpclcd, manuscript = letter]{achemso}
\usepackage{achemso}
\usepackage{amsmath}
\usepackage{amssymb}
\usepackage{graphicx}
\usepackage{color}
\usepackage[group-separator={,},group-minimum-digits=4]{siunitx}
\usepackage{subfig}
\title{Reproducing Quantum Probability Distributions at the Speed of Classical Dynamics: A New Approach for Developing Force-Field Functors}

\author{Vikram Sundar}
\affiliation{Department of Chemistry and Chemical Biology, Harvard University, Cambridge, MA 02138}
\email{vikramsundar@college.harvard.edu}

\author{David Gelbwaser-Klimovsky}
\affiliation{Department of Chemistry and Chemical Biology, Harvard University, Cambridge, MA 02138}
\email{dgelbwaser@fas.harvard.edu}

\author{Al\'an Aspuru-Guzik}
\affiliation{Department of Chemistry and Chemical Biology, Harvard University, Cambridge, MA 02138}
\alsoaffiliation{Senior Fellow, Canadian Institute for Advanced Research, Toronto, Ontario M5G 1Z8, Canada} 
\email{alan@aspuru.com}

\begin{document}

\begin{tocentry}
  \includegraphics[width=\textwidth]{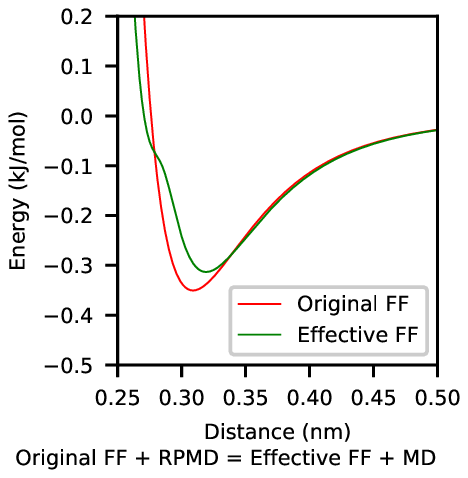}
\end{tocentry}

\begin{abstract}
Modeling nuclear quantum effects is required for accurate molecular dynamics (MD) simulations of molecules. The community has paid special attention to water and other biomolecules that show hydrogen bonding. Standard methods of modeling nuclear quantum effects like Ring Polymer Molecular Dynamics (RPMD) are computationally costlier than running classical trajectories. A force-field functor (FFF) is an alternative method that computes an effective force field which replicates quantum properties of the original force field. In this work, we propose an efficient method of computing FFF using the Wigner-Kirkwood expansion. As a test case, we calculate a range of thermodynamic properties of Neon, obtaining the same level of accuracy as RPMD, but with the shorter runtime of classical simulations. By modifying existing MD programs, the proposed method could be used in the future to increase the efficiency and accuracy of MD simulations involving water and proteins. 
\end{abstract}

\maketitle


Biomolecular simulation methods have proven increasingly useful for understanding properties of biomolecules, such as protein binding and protein folding \cite{simulation1, simulation2}. The most popular simulation methods today are molecular dynamics (MD) and Monte Carlo (MC) simulations, both of which rely on empirical force fields to model interactions between molecules \cite{simulation1, simulation2, forceField3}. Most modern force fields are developed by fitting functional forms to the Born-Oppenheimer energy surface of the molecule \cite{forceField1, forceField2, forceField3, allentildesley}. Afterwards, Monte Carlo and molecular dynamics are used to simulate atoms as classical particles moving on this Born-Oppenheimer surface \cite{allentildesley}, ignoring quantum effects on the nuclear dynamics.

Accounting for nuclear quantum properties is essential for accurate biomolecular simulations using either molecular dynamics or Monte Carlo methods since they play a role in the thermodynamic properties of water and other fluids \cite{nuclearImp, nuclearImp4}, hydrogen-bonding \cite{nuclearImp4} and the binding of enzymes to ligands \cite{nuclearImp2, nuclearImp3}. Since Born-Oppenheimer force fields and classical MD do not account for these effects, a number of alternative methods have been
developed to account for them. These include Path Integral Molecular Dynamics (PIMD) \cite{allentildesley, oldNuclear10, oldNuclear11Both, oldNuclear8}, Ring-Polymer Molecular Dynamics (RPMD) \cite{oldNuclear12}, and Centroid Molecular Dynamics (CMD) \cite{oldNuclear3}, the generalized Langevin equation \cite{oldNuclear1}, higher-order factorization schemes \cite{oldNuclear7, oldNuclear9}, and other methods \cite{oldNuclear4Both, oldNuclear5}. These methods have been used in simulations
\cite{oldnuclear2Sim, oldNuclear4Both, oldNuclear6Sim} but are slower than classical MD since they require to increase the number of particles by a factor of 32 in order to achieve sufficient accuracy \cite{rpmd}.

In this letter, we put forward an alternative method for performing quantum simulations at the speed of classical simulations with no loss in accuracy. In \citet{functor}, we proved the existence and uniqueness of a force-field functor (FFF) that computes an effective potential which accounts for nuclear quantum corrections and can be used in classical simulations to calculate thermodynamic properties. Specifically, we defined the map $\mathcal{F}$ from a classical potential $V(q)$ with equilibrium quantum density $n_Q (q)$ to the effective potential $W(q)$ with equilibrium classical density $n_0 (q) = n_Q (q)$. The path integral formulation of quantum mechanics allows us to explicitly write \begin{equation}
  W(q) = - \frac{1}{\beta} \log \left[\oint_{r(0) = q}^{r(\beta \hbar) = q} \mathcal{D}r (\tau) \, e^{- A[r(\tau)]}\right]
  \label{eq:Functor}
\end{equation} with Boltzmann factor $\beta$ and Helmholtz free energy $A$. In \citet{functor}, we demonstrated the existence of this map and showed its uniqueness. 

$W(q)$ can be used to compute the equilibrium particle density and partition function, and thus all thermodynamic properties including nuclear quantum effects, at a cost equivalent to classical dynamics \cite{functor}. Further, uniqueness of the FFF implies that any method of computing an effective potential that has the correct partition function and equilibrium density will satisfy these properties. We propose a method of computing the effective potential using the Wigner-Kirkwood expansion \cite{wigner, kirkwood} which was originally introduced to compute quantum corrections of thermal states. This expansion, when taken to infinite order, correctly reproduces the quantum partition function, so it must be the correct effective potential \cite{wigner, kirkwood}.

\citet{wigner} introduced the Wigner function to compute quantum corrections to a thermal state.
The Wigner function of a thermal state of a particle in a potential $V(\vec{r})$ can be expanded as a function of $\hbar$


\begin{equation}
  W(\vec{r}, \vec{p}) = e^{-\beta \mathcal{H}(\vec{r}, \vec{p})}\left(1 + \beta^2 \hbar^2 W_2 (\vec{r}, \vec{p}) + \beta^4 \hbar^4 W_4 (\vec{r}, \vec{p}) + \ldots\right),
  \label{eqn:WignerExpansion}
\end{equation}

\noindent where $\mathcal{H}(\vec{r}, \vec{p})$ is the Hamiltonian and $\beta = \frac{1}{k_b T}$. 
The $W_i (\vec{r}, \vec{p})$ are functions of derivatives of the potential function $V(\vec{r})$; methods to compute them can be found in \citet{wigner} and \citet{wignerCompute} 
The Wigner function can be used to compute the probability distribution in position by integrating out the momentum coordinate. The effective potential, $\widetilde{V}(\vec{r})$, that reproduces this distribution in a classical thermal state is then 




\begin{equation}
  \widetilde{V}(\vec{r}) = - \frac{1}{\beta} \log \left[\int d\vec{p} \, W(\vec{r}, \vec{p})\right].
  \label{eqn:FunctorDef}
\end{equation}







At high enough temperatures, any system behaves classically and the Wigner function is equal to the classical phase space distribution, the first term in the Wigner expansion. At intermediate temperatures, quantum effects are relevant and are fully described by the first correction to the classical phase-space distribution, $W_2$ \cite{wigner}. For a spherically symmetric system, the non-normalized correction term is


\begin{equation}
  W_2 (r, p, \theta)  = \frac{1}{24 m^2 r} \left(-6 m \frac{\partial V}{\partial r} + \beta p^2 \sin^2 \theta \frac{\partial V}{\partial r} + m r \beta \left(\frac{\partial V}{\partial r}\right)^2 - 3 m r \frac{\partial^2 V}{\partial r^2} + \beta p^2 r \cos^2 \theta \frac{\partial^2 V}{\partial r^2}\right).
  \label{eqn:Wigner2OrderSpherical}
\end{equation}

\noindent Here, $r$ is radial distance, $p$ is the magnitude of the momentum, and $\theta$ is the angle between position and momentum. 

For our analysis to hold, the force field must be fitted to high-quality electronic structure data. This will ensure that the initial classical force field accurately models dynamics according to the electronic structure. We can then use the Born-Oppenheimer approximation to isolate nuclear and electronic effects and use the FFF to add nuclear quantum effects. Many force fields are fitted to experimental data like the radial distribution function, heat of vaporization, or density which already account for nuclear quantum effects; using the FFF or other quantum simulation methods on these force fields would double-count those effects \cite{doubleCount}. However, these force fields have generally had limited transferability in measuring experimental parameters that they are not fitted to.

Previous efforts to use Wigner-Kirkwood expansions to add quantum corrections to MD calculations have had mixed results. Instead of using an effective potential, many of these efforts\cite{allentildesley, oldThermo, oldThermo2, oldThermo3, oldThermo4, oldThermo5, oldThermo6} added quantum corrections to thermodynamic quantities after the simulation. This approach is more time-consuming, since corrections need to be computed for every thermodynamic quantity of interest \cite{allentildesley}. \citet{oldEffective} used an effective potential on a Neon force field fitted to low-quality electronic structure data, so their results, in particular the position of the radial distribution function of nearest-neighbor peaks, did not show an unambiguous improvement over the classical potential. We demonstrate that an effective potential derived from high-quality \emph{ab initio} data can be used to calculate condensed-phase properties with high accuracy and significant improvements over classical calculations. 


We test our method on a Neon force field. Neon's interatomic potential is spherically symmetric and a highly accurate force field for Neon was derived by \citet{NeonFF} by fitting functional forms to high-quality CCSD(T) electronic structure data.

\begin{figure}[!htb]
  \centering
  \includegraphics[width=\textwidth]{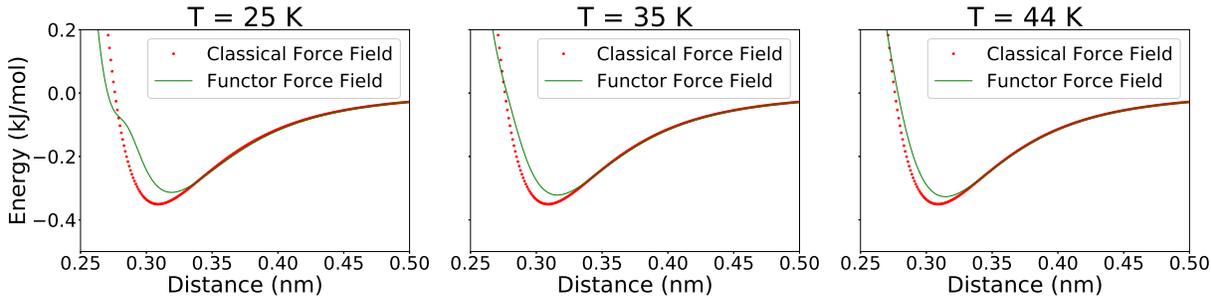}
  \caption{Classical and effective force field for Neon at $T = \SI{25}{\kelvin}, \SI{35}{\kelvin},$ and $T = \SI{44}{\kelvin}$. The red dotted curve is the classical potential from \citet{NeonFF} and the green curve is the effective potential numerically computed using the Wigner expansion to second order. The change in shape and shallower well can be explained by the wavepacket entering the classically forbidden region and the zero-point energy, respectively.}
  \label{fig:functorPotential}
\end{figure}

We began by numerically computing the effective potential at various temperatures as shown in Figure \ref{fig:functorPotential}. We note two characteristics of the effective potential which can be explained by quantum phenomena: First, the repulsive wall as $r \to 0$ is less steep in the effective potential than in the classical potential. This is due to the fact that the quantum wavepacket can enter the classically forbidden region $E < V(r)$. Second, the well is shallower, due to the zero-point energy.

Moreover, the equilibrium distance is slightly longer in the effective potential than in the classical potential. As we show below, this has important thermodynamic consequences.

We note a bump in the effective force field at $T = \SI{25}{\kelvin}$. This is likely an artifact due to use of only the second-order expansion, and not a reflection of the full quantum effective force field. Addition of subsequent terms in the Wigner expansion should eliminate this bump \cite{oldThermo5}.

\begin{figure}[!htb]
  \centering
  \includegraphics[width=\textwidth]{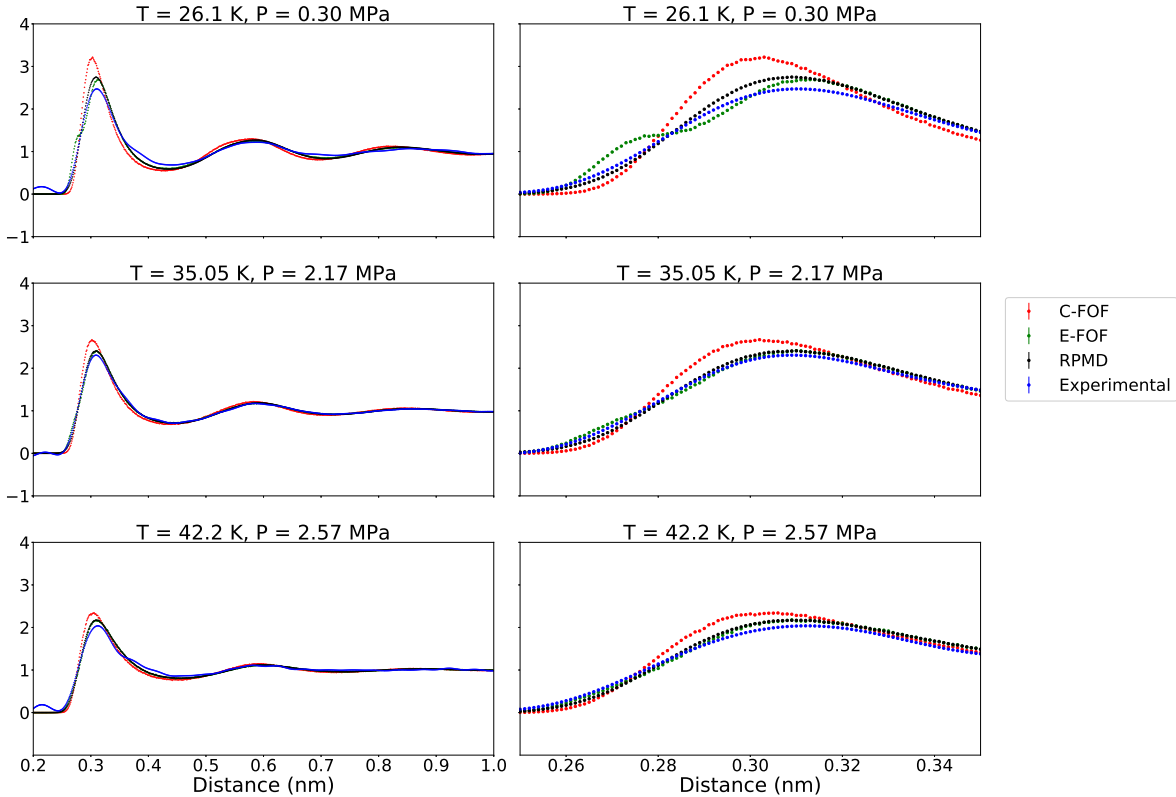}
  \caption{Radial distribution function as computed by C-FOF, E-FOF, and RPMD. In each plot, the red curve is computed by C-FOF, the green curve is computed by E-FOF, the black curve is computed by RPMD, and the blue curve is experimentally determined from \citet{NeonRDF26} and \citet{NeonRDF35}. The right side focuses on just the first peak. In all cases, the height and location of the first peak are more accurately determined by E-FOF and RPMD than by C-FOF. E-FOF demonstrates similar accuracy as the significantly slower RPMD.}
  \label{fig:RDF}
\end{figure}

Figure \ref{fig:RDF} shows radial distribution functions (RDF) for liquid Neon at $T = \SI{26.1}{\kelvin}$, $T = \SI{35.05}{\kelvin}$, and $T = \SI{42.2}{\kelvin}$. Experimental data \cite{NeonRDF26, NeonRDF35} is compared to three simulation methods: a classical MD simulation with the classical force field (henceforth called C-FOF); a standard RPMD simulation with the classical force field which accounts for nuclear quantum effects (henceforth called RPMD); and a classical MD simulation with the effective force field (henceforth called E-FOF). 

At $T = \SI{35.05}{\kelvin}$ and $T = \SI{42.2}{\kelvin}$, E-FOF is roughly as accurate as RPMD, with both being a significant improvement relative to C-FOF. In particular, the location and height of the first peak obtained by E-FOF are more precise than those derived by C-FOF. The shift and lower height of the peak are caused by quantum corrections which shift the equilibrium position and reduce the depth of the well in the effective potential described above.

At $T = \SI{26.1}{\kelvin}$, E-FOF, while still a significant improvement over C-FOF, is somewhat worse than RPMD. Specifically, the location and height of the peak are relatively accurate, but the bump in the potential noted above results in a similar bump in the RDF. Inaccuracies in both RPMD and E-FOF at this temperature indicate more beads and higher-order terms are necessary for an accurate computation of the RDF. 

We also measured the location of the first peak of the radial distribution function and compare our results with those of \citet{oldEffective} in Table \ref{tab:RDFErmakova}. As noted earlier, we demonstrate clear, consistent improvement in our computed radial distribution function from C-FOF to E-FOF and moderate improvement over the results computed by \citet{oldEffective}. This emphasizes the need of high-quality electronic data in order to obtain accurate radial distribution functions.

RPMD was around $100$ times slower on similar-capability machines than E-FOF. This is expected since RPMD had $P = 32$ beads per atom and MD simulations scale as $\mathcal{O}(N^2)$ where $N$ is the number of particles simulated. Thus E-FOF reproduces the results of quantum simulations at the significantly faster speed of classical simulations. Computing speed is especially relevant for more complex calculations such as the equation of state and vapor-liquid coexistence curve.

\begin{table}
  \centering
  \begin{tabular}{c | c | c | c}
    Temperature & $\SI{26}{\kelvin}$ & $\SI{35}{\kelvin}$ & $\SI{42}{\kelvin}$ \\
    \hline
    Our C-FOF & $\SI{303}{\pico\meter}$ & $\SI{302}{\pico\meter}$ & $\SI{306}{\pico\meter}$ \\
    Our E-FOF & $\SI{313}{\pico\meter}$ & $\SI{310}{\pico\meter}$ & $\SI{309}{\pico\meter}$ \\
    \citet{oldEffective} C-FOF & $\SI{306}{\pico\meter}$ & $\SI{307}{\pico\meter}$ & $\SI{308}{\pico\meter}$ \\
    \citet{oldEffective} E-FOF & $\SI{314}{\pico\meter}$ & $\SI{312}{\pico\meter}$ & $\SI{312}{\pico\meter}$ \\
    Experimental & $\SI{310}{\pico\meter}$ & $\SI{310}{\pico\meter}$ & $\SI{310}{\pico\meter}$
  \end{tabular}
  \caption{Position of the first peak of the RDF. Our results (E-FOF) are closer to the experimental data than C-FOF and previous results \cite{oldEffective}.}
  \label{tab:RDFErmakova}
\end{table}


\begin{figure}[!htb]
  \centering
  \includegraphics[width=0.5\textwidth]{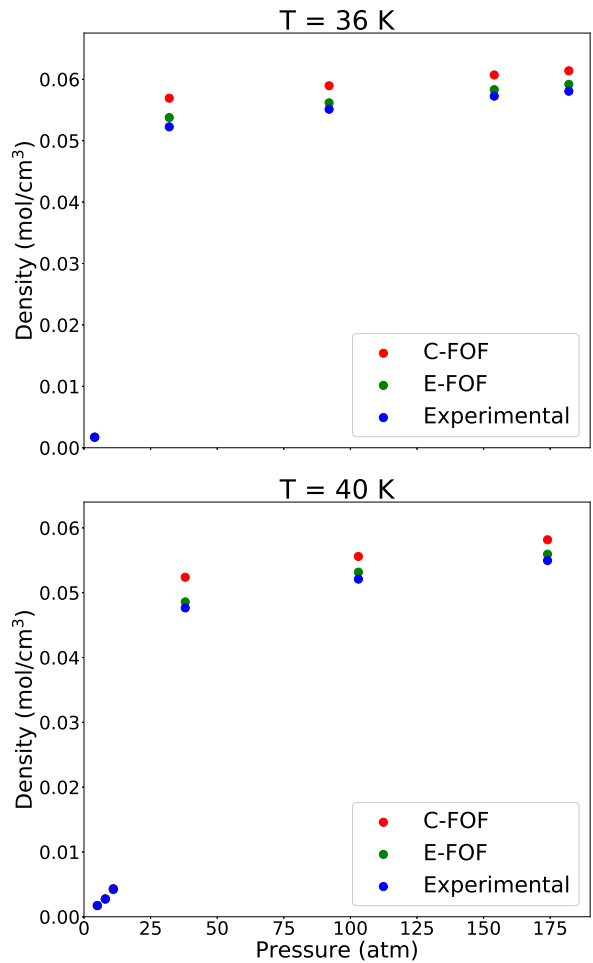}
  \caption{Equation of state as computed by C-FOF and E-FOF at $T = \SI{36}{\kelvin}, \SI{40}{\kelvin}$. The red points were computed by C-FOF, the green points by E-FOF, and the blue points experimentally determined by \citet{NeonDensity}. The gas-phase density points from all three methods are too close to be distinguished. E-FOF is significantly more accurate in liquid-phase, reducing the error relative to experimental data from $10\%$ for C-FOF to $2\%$ for E-FOF.}
  \label{fig:EoS}
\end{figure}

Figure \ref{fig:EoS} shows the equation of state for Neon at $T = \SI{36}{\kelvin}$ and $T =  \SI{40}{\kelvin}$. Experimental data \cite{NeonDensity} is compared to densities from NPT MD simulations for both the classical and effective potential \cite{openmm}. These values span liquid and gas phase. 

Both E-FOF and C-FOF were accurate on gas-phase densities. The maximum error relative to the experimental data for C-FOF is around $1\%$ and for E-FOF around $2\%$. This is somewhat expected since gas-phase densities do not depend much on the force field; we see that the ideal gas term (independent of potential) dominates in the equation of state relative to the virial coefficients (which depend on potential). In liquid phase, C-FOF significantly overestimates the experimental density by as much as $10\%$ while E-FOF only overestimates the experimental density by around $2\%$. This is due to the shift in equilibrium position described above. Adding additional terms to the Wigner expansion should further reduce the error in calculated density.


\begin{figure*}[!htb]
  \centering
  \subfloat[Density vs. Temperature]{\includegraphics[width=0.5\textwidth]{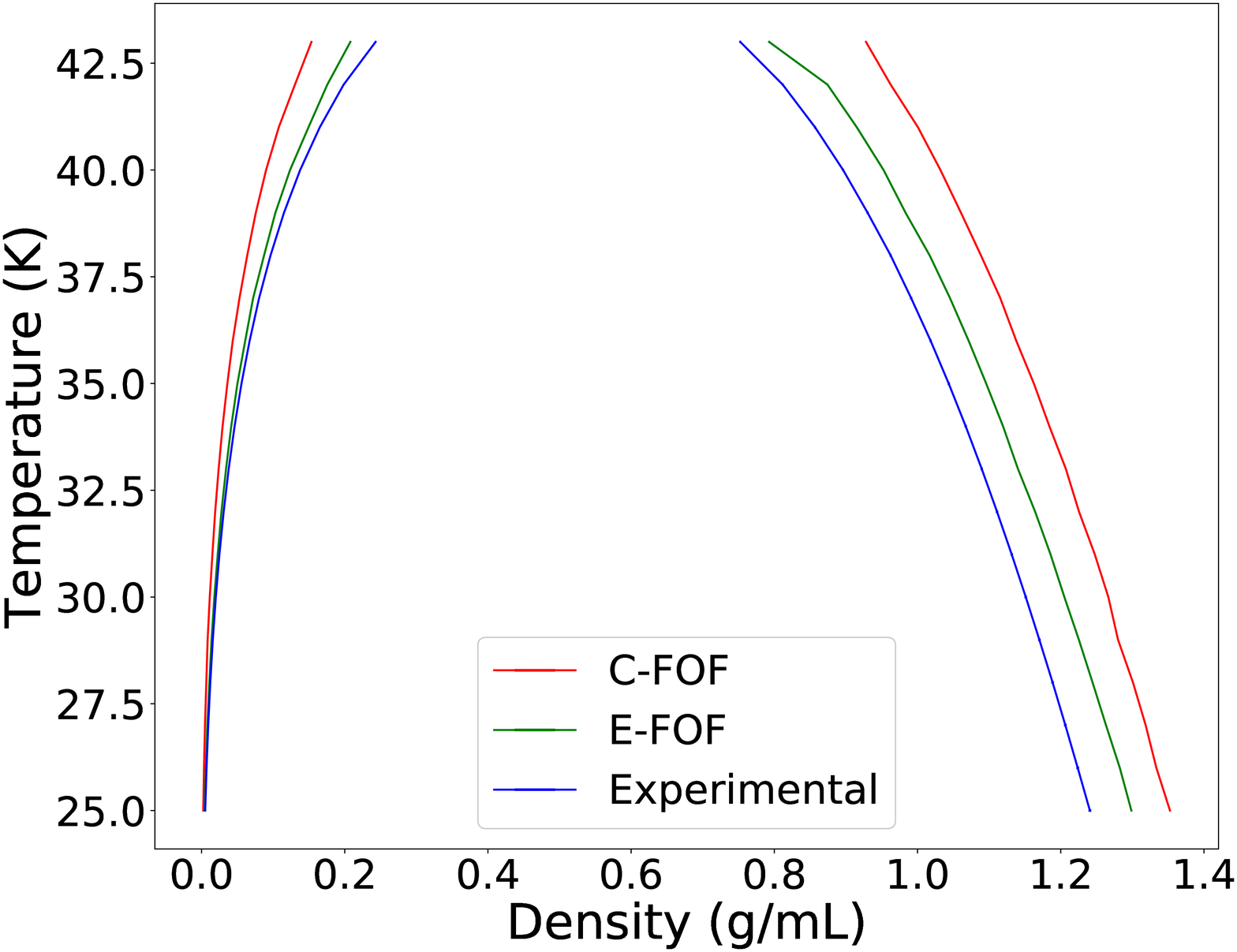}}
  \subfloat[Temperature vs. Pressure]{\includegraphics[width=0.5\textwidth]{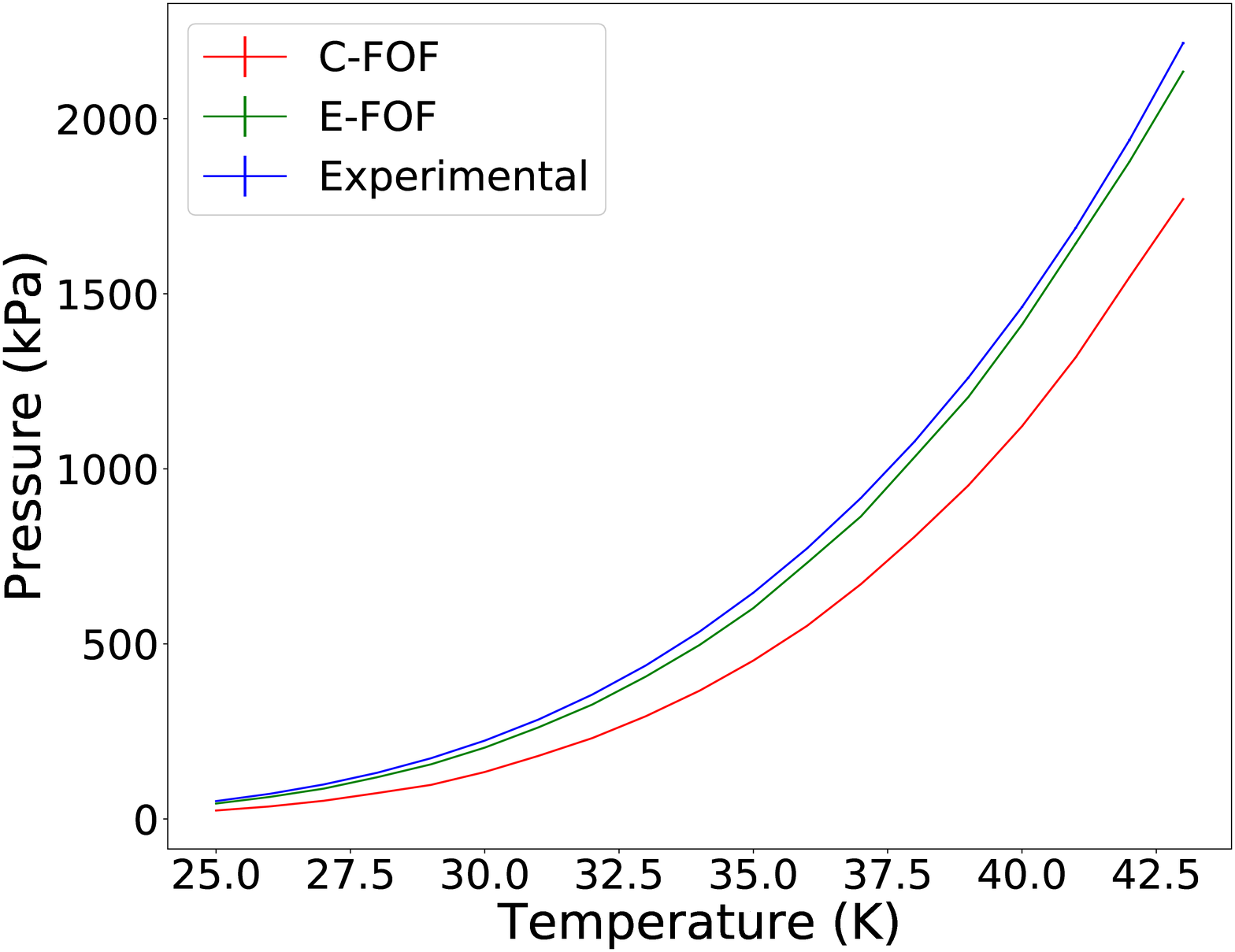}}
  \hfill
  \subfloat[Absolute Density Error vs. Density]{\includegraphics[width=0.5\textwidth]{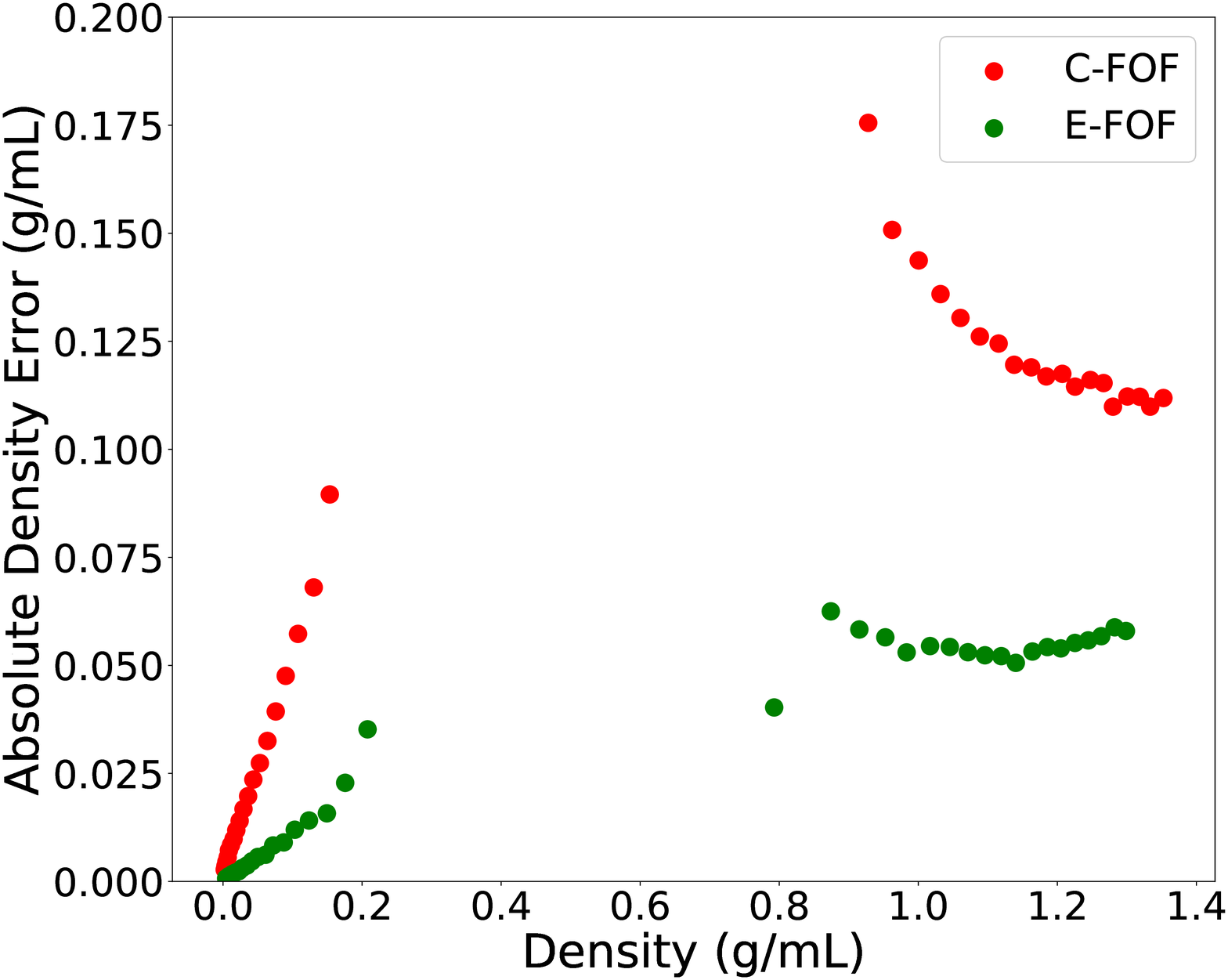}}
  \subfloat[Absolute Pressure Error vs. Temperature]{\includegraphics[width=0.5\textwidth]{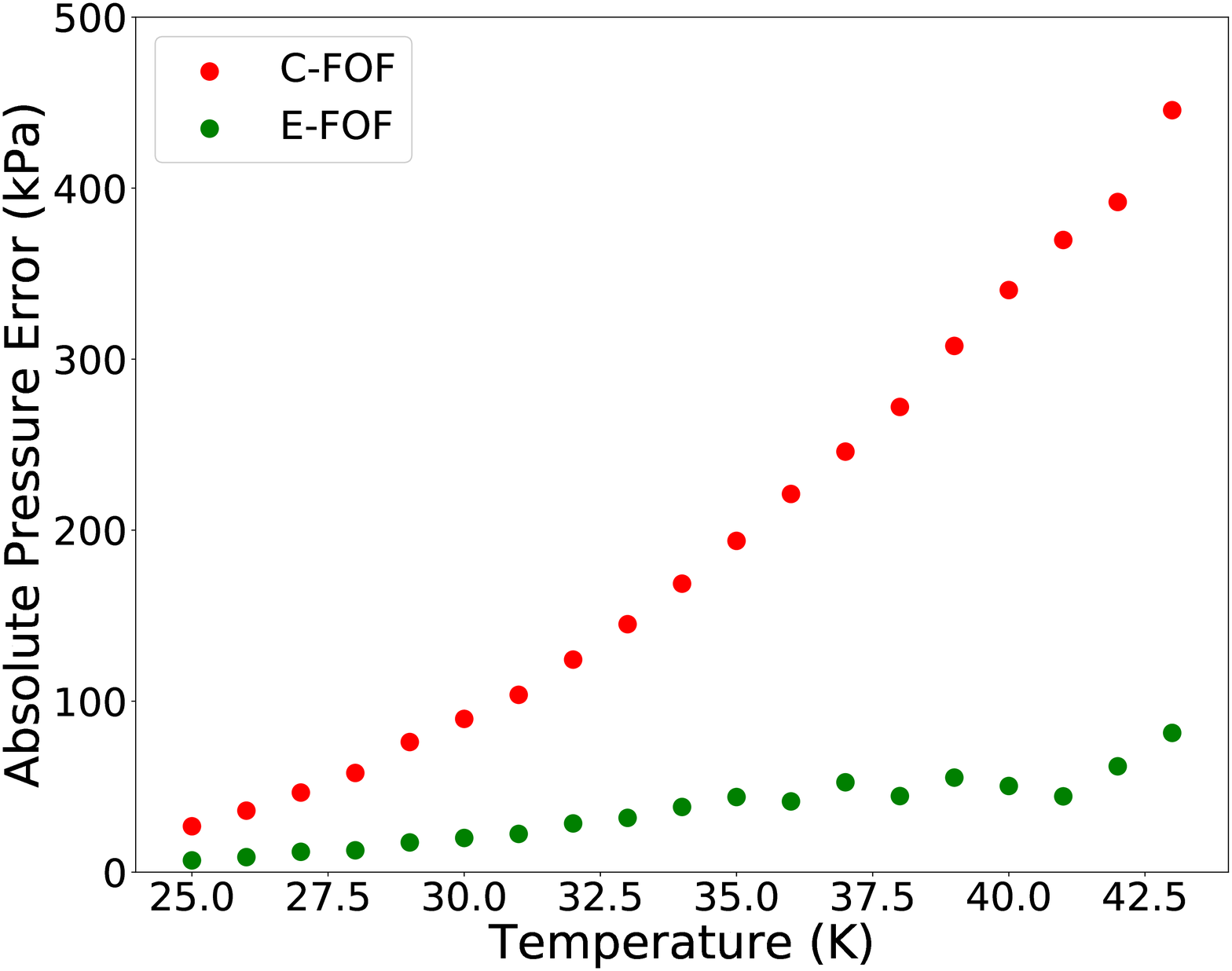}}
  \caption{Vapor-liquid coexistence curve for Neon, as computed by Gibbs MC simulation with both C-FOF and E-FOF. The red curves are computed by C-FOF, the green curves by E-FOF, and the blue curves experimentally determined by \citet{NeonBPdata}. E-FOF is significantly more accurate in all cases. It reduces the average error relative to the experimental data: from $40\%$ to $11\%$ for the gas phase density; from $12\%$ to $5\%$ for the liquid phase density; from $33\%$ to $7\%$ for the pressure.}
  \label{fig:NeonBP}
\end{figure*}

We computed the vapor-liquid coexistence curve with Gibbs Monte Carlo simulations for both the classical and effective potential \cite{gibbs1, gibbs2, gibbs3, towhee}. Experimental data for both densities and pressures was obtained from \citet{NeonBPdata} We ran simulations ranging from close to the triple point to close to the critical point, thus covering the entire range of the vapor-liquid coexistence curve.


E-FOF is significantly more accurate in determining both the density and the pressure of the vapor-liquid coexistence point, as shown in Figure \ref{fig:NeonBP}. Specifically, C-FOF consistently overestimates the liquid density by an average percent error of $12\%$ relative to experimental data, while E-FOF consistently overestimates it by only $5\%$. This effect can be partly explained by the density deviations observed when computing the equation of state. C-FOF underestimates gas density by an average percent error of $40\%$ relative to the experimental data, while E-FOF underestimates it by $11\%$. Finally, C-FOF underestimates the pressure by an average percent error of $33\%$ relative to the experimental data while E-FOF underestimates it by $7\%$. Larger errors at lower temperature suggest that the effective force field would be improved by using higher-order corrections (see Equation \ref{eqn:WignerExpansion}).

We also note that E-FOF began exhibiting critical behavior, i.e., transitions  between vapor and liquid phase, at around $T = \SI{43}{\kelvin}$, very close to the experimental critical point of $T_c = \SI{44.40}{\kelvin}$. C-FOF did not exhibit similar behavior even at $T = \SI{44}{\kelvin}$. We chose not to measure the critical point directly due to inaccuracies with Gibbs MC simulations close to the critical point \cite{gibbsCrit}, but this result suggests that C-FOF also overestimates the critical point.


Our results conclusively show that the effective force field is significantly more accurate than the classical force field for Neon on a wide range of condensed-phase parameters. We have also demonstrated that the effective force field was capable of computing radial distribution functions to the same degree of accuracy as quantum simulations with the speed of a classical simulation, i.e., the FFF can be used to account for nuclear quantum effects about a hundred times faster than standard methods. The FFF can be applied to any force field which is fitted to sufficiently high-quality quantum energy data and for whom subsequent terms in the Wigner expansion can reasonably be suppressed to accurately account for nuclear quantum effects at the speed of a classical simulation.

Next steps include extending this method to non-spherical force fields and force fields for biomolecules, which are relevant to biochemists. The FFF could significantly correct simulations for any biomolecule involving hydrogen bonding or hydrogen atoms in a central way, including water, proteins, and nucleic acids \cite{nuclearImp, nuclearImp2, nuclearImp3, nuclearImp4}. Applying the FFF to standard protein force fields like AMBER and CHARMM \cite{charmm, amber} should result in accurate quantum simulations of biomolecules with no additional computational cost. 



This project was partially funded by a grant from the Harvard College Research Program, a grant from the Harvard University Center for the Environment Undergraduate Research Fund, and a grant from the Harvard Physics Department. D. G.-K. and A. A.-G. acknowledge the support from the Center for Excitonics, an Energy Frontier Research Center funded by U.S. Department of Energy under award DE-SC0001088. Numerical simulations for this project were run on the Harvard Odyssey cluster.



\begin{suppinfo}

Details of simulation methods can be found in the supplement.

\end{suppinfo}

\bibliography{Neon}

\end{document}